\documentclass{article}
\usepackage{graphicx}
\usepackage{amsmath}
\usepackage{amsfonts}
\usepackage{amssymb}
\begin{document}

\begin{center}
{\Large Nonlinear quantum mechanics implies polynomial-time solution for
NP-complete and \#P problems}\footnote{This work supported in part by grant \#
N00014-95-1-0975 from the Office of Naval Research, and by ARO and DARPA under
grant \# DAAH04-96-1-0386 to QUIC, the Quantum Information and Computation
initiative, and by a DARPA grant to NMRQC, the Nuclear Magnetic Resonance
Quantum Computing initiative.}

\bigskip\medskip

Daniel S. Abrams

Department of Physics, MIT 12-128b

Cambridge, MA 02139 (abrams@mit.edu)

\medskip\medskip

Seth Lloyd

d'Arbeloff Laboratory for Information Sciences and Technology

Department of Mechanical Engineering, MIT 3-160

Cambridge, MA 02139 (slloyd@mit.edu)

\medskip\medskip
\end{center}

If quantum states exhibit small nonlinearities during time evolution, then
quantum computers can be used to solve NP-complete problems in polynomial
time. We provide algorithms that solve NP-complete and \#P oracle problems by
exploiting nonlinear quantum logic gates. It is argued that virtually any
deterministic nonlinear quantum theory will include such gates, and the method
is explicitly demonstrated using the Weinberg model of nonlinear quantum
mechanics.\bigskip\bigskip\bigskip\pagebreak 

Computers are physical devices: like all other physical systems, their
behavior is determined by physical laws. This seemingly obvious statement has
important implications, because it indicates that as our understanding of
physical phenomena expands, the theoretical limits to the power of computing
machines may grow accordingly. Recently, it has been shown that quantum
computers can in theory exploit quantum phenomena to perform tasks that
classical computers apparently cannot, such as factoring large numbers in
polynomial time\cite{Shor}, searching databases of size $M$ in time $\sqrt{M}$
\cite{Grover}, or simulating the detailed behavior of other quantum systems in
less than exponential time and space \cite{Feynman}\cite{Lloyd}\cite{Abrams}.
The realization that quantum mechanics could be used to build a fundamentally
more powerful type of computing machine has led to a huge amount of recent
activity in the field of quantum computation; for a review, see
Ekert\cite{Ekert} or DiVincenzo\cite{DiVincenzo}.

It has been suggested \cite{Weinberg}\cite{Weinberg 2}\cite{Fivel}\cite{Levy}
\cite{Bertolami} that under some circumstances the superposition principle of
quantum mechanics might be violated - that is, that the time evolution of
quantum systems might be (slightly) nonlinear. Such nonlinearity is purely
hypothetical: all known experiments confirm the linearity of quantum mechanics
to a high degree of accuracy\cite{Majumder}\cite{Walsworth}\cite{Chupp}%
\cite{Bollinger}. Further, nonlinear quantum theories have often been
controversial and frequently have had theoretical difficulties\cite{Peres}
\cite{Polchinski}\cite{Gisin}. Nevertheless, the implications of nonlinear
quantum mechanics on the theory of computation are profound. In particular, we
show that it is generally possible to exploit nonlinear time evolution so that
the classes of problems NP and \#P\ (including oracle problems) may be solved
in polynomial time. An experimental question - that is, the exact linearity of
quantum mechanics - could thereby determine the answer to what may have
previously appeared to be a purely mathematical one. This paper therefore
establishes a new link between physical law and the theoretical power of
computing machines.

The class NP is the set of problems for which, once given a potential answer,
one can determine in polynomial time if the potential answer is in fact a
solution. These include all problems in the class P\ (those that can be solved
in polynomial time) as well as the NP-complete problems, e.g., the traveling
salesman, satisfiability, and sub-graph isomorphism, for which no known
polynomial time algorithms exist. We phrase our algorithm in terms of an
oracle (or ``black box''), which calculates a function that maps n bits into a
single bit (i.e., it takes an input between 0 and $2^{n}-1$ and returns either
0 or 1). We need to determine if there exists an input value x for which f(x)
= 1; with a polynomial time algorithm to solve this problem, it is easy to
solve all problems in the class NP. Indeed, this oracle problem is in fact a
harder problem than those in NP, because it clearly requires exponential time
on a classical Turing machine.

A simple algorithm that solves the NP oracle problem can be thought of as an
extension of Grover's data-base search algorithm\cite{Grover} to a nonlinear
regime. Suppose that it is possible to perform a nonlinear operation on a
single qubit that has a positive Lyapunov exponent over some (not
exponentially small) region: i.e., somewhere on the unit sphere there exists a
line of finite extent along which application of the operation causes nearby
points to move apart exponentially at a rate $e^{\lambda\Delta\theta}$
proportional to their original separation $\Delta\theta.$ We can exploit this
behavior to solve NP problems if we begin with an ordinary quantum computer
(i.e., one that can perform the usual quantum logic operations such as
controlled rotation gates) and use the algorithm described below:

Step 1. Begin by performing a $\pi/2$ rotation on each of the first n qubits
to obtain the state%

\begin{equation}
\psi=\frac{1}{\sqrt{2^{n}}}\sum\limits_{i=0}^{2^{n}-1}|i,0\rangle
\end{equation}

Step 2. Use the oracle (only once) to calculate f(i):%

\begin{equation}
\psi=\frac{1}{\sqrt{2^{n}}}\sum\limits_{i=0}^{2^{n}-1}|i,f(i)\rangle
\end{equation}

Step 3. Reverse the $\pi/2$ rotation on each of the first n qubits. Each
state
%TCIMACRO{\TEXTsymbol{\vert}}%
%BeginExpansion
$\vert$%
%EndExpansion
i%
%TCIMACRO{\TEXTsymbol{>}}%
%BeginExpansion
$>$%
%EndExpansion
then maps into a superposition over all possible
%TCIMACRO{\TEXTsymbol{\vert}}%
%BeginExpansion
$\vert$%
%EndExpansion
i%
%TCIMACRO{\TEXTsymbol{>}}%
%BeginExpansion
$>$%
%EndExpansion
, with amplitude $\pm\frac{1}{\sqrt{2^{n}}}$. In particular, each state
%TCIMACRO{\TEXTsymbol{\vert}}%
%BeginExpansion
$\vert$%
%EndExpansion
i%
%TCIMACRO{\TEXTsymbol{>}}%
%BeginExpansion
$>$%
%EndExpansion
contributes $+\frac{1}{\sqrt{2^{n}}}$of its amplitude to the state
%TCIMACRO{\TEXTsymbol{\vert}}%
%BeginExpansion
$\vert$%
%EndExpansion
00...0%
%TCIMACRO{\TEXTsymbol{>}}%
%BeginExpansion
$>$%
%EndExpansion
, for a total contribution of amplitude $\frac{1}{2^{n}}$ from each
%TCIMACRO{\TEXTsymbol{\vert}}%
%BeginExpansion
$\vert$%
%EndExpansion
i%
%TCIMACRO{\TEXTsymbol{>}}%
%BeginExpansion
$>$%
%EndExpansion
. At least $\frac{1}{2}2^{n}$ of these states correspond to a particular value
of f(i)=a, and thus the state
%TCIMACRO{\TEXTsymbol{\vert}}%
%BeginExpansion
$\vert$%
%EndExpansion
00....0,a%
%TCIMACRO{\TEXTsymbol{>}}%
%BeginExpansion
$>$%
%EndExpansion
has amplitude at least 1/2. A measurement on the first n qubits will therefore
yield the state
%TCIMACRO{\TEXTsymbol{\vert}}%
%BeginExpansion
$\vert$%
%EndExpansion
00...0%
%TCIMACRO{\TEXTsymbol{>}}%
%BeginExpansion
$>$%
%EndExpansion
with probability at least 1/4. The system will then be in the state%

\begin{equation}
\psi=\frac{2^{n}}{\sqrt{2^{2n}-2^{n+1}s+2s^{2}}}\left|  00..0\right\rangle
\otimes\left\{  \frac{2^{n}-s}{2^{n}}\left|  0\right\rangle +\frac{s}{2^{n}%
}\left|  1\right\rangle \right\}
\end{equation}

where s is the number of solutions i for which f(i)=1.\ The last qubit now
contains all the information that we need; however, for small s, a measurement
of the last qubit will almost always return
%TCIMACRO{\TEXTsymbol{\vert}}%
%BeginExpansion
$\vert$%
%EndExpansion
0%
%TCIMACRO{\TEXTsymbol{>}}%
%BeginExpansion
$>$%
%EndExpansion
, yielding no information. We wish to distinguish between the cases s=0 and s%
%TCIMACRO{\TEXTsymbol{>}}%
%BeginExpansion
$>$%
%EndExpansion
0.

Step 4. Repeatedly apply the nonlinear operation to drive the states
representing these two cases apart at an exponential rate:\ eventually, at a
time determined by a polynomial function of the number of qubits n, the number
of solutions s, and the rate of spreading (Lyapunov exponent) $\lambda$, the
two cases will become macroscopically distinguishable.

Step 5. Make a measurement on the last qubit to determine the solution. If the
angular extent of the nonlinear region is small, it may be necessary to repeat
the algorithm several times in order to determine the solution with high
probability. In general, the algorithm will require $O((\pi/\eta)^{2})$
trials, where $\eta$ is the angular extent of the nonlinear region. If $\eta$
is sufficiently large, the oracle may need to be called only once.

To solve problems in the class \#P, we need to determine the exact number of
solutions s. This is approximately found by counting the number of times that
the nonlinear operator was applied. To determine s exactly, one proceeds with
finer and finer estimates by rotating the final qubit such that the current
best estimate is centered in the nonlinear region; in this way, applying the
nonlinear operator separates states with s near this value so that they are
distinguishable. With only a polynomial number of iterations, one determines
the value s exactly. The previous algorithm can therefore by used to solve
arbitrary problems (including oracle probems) in the class \#P\ as well.

The above algorithm has one disadvantage in that it requires exponential
precision.\ We describe below another algorithm, which (at least in the case
of the Weinberg model) appears to be robust against small errors. In order to
simplify the description, we assume (for now) that there is at most a single
value x for which f(x) = 1. We begin as before with a quantum computer that
can perform the usual quantum logic operations, and that can in addition
perform a simple nonlinear operator whose form will be described below. We
conceptually divide the qubits into $n$ input qubits (representing the input,
an integer between 0 and $2^{n}-1$ ) and one flag qubit. In order to solve the
stated oracle problem, we use the following algorithm:

Step 1. We begin by performing a $\pi/2$ rotation on each of the first n
qubits to obtain the state%

\begin{equation}
\psi=\frac{1}{\sqrt{2^{n}}}\sum\limits_{i=0}^{2^{n}-1}|i,0\rangle
\end{equation}

Step 2. Use the first N qubits as the input to the oracle and store the output
in the\ (N+1)$^{st}$ qubit (the flag qubit). (Note that this is the only time
that the algorithm needs to call the oracle; hence, we require only 1
evaluation of the function.) The system is now in the state%

\begin{equation}
\psi=\frac{1}{\sqrt{2^{n}}}\sum\limits_{i=0}^{2^{n}-1}|i,f(i)\rangle
\end{equation}

Step 3. Consider the first qubit separately, and group all the states of the
superposition into pairs based on the value of qubits 2..n. That is, the
qubits 2..n define $2^{n-1}$ subspaces of dimension 4 = (2 dimensions for
qubit 1) * (2 dimensions for the flag qubit). Within each subspace, the
computer will be in one of the following states (where we write the value of
the first qubit followed by the value of the flag qubit, and ignore the
normalization constants):%

\begin{align}
&  (a)\;\;|00\rangle+|11\rangle\nonumber\\
&  (b)\;\;|01\rangle+|10\rangle\\
&  (c)\;\;|00\rangle+|10\rangle\nonumber
\end{align}

(At the start of the computation, most of the superposition will be in the
third state, because the flag qubit is
%TCIMACRO{\TEXTsymbol{\vert}}%
%BeginExpansion
$\vert$%
%EndExpansion
1%
%TCIMACRO{\TEXTsymbol{>}}%
%BeginExpansion
$>$%
%EndExpansion
in at most only 1 of the $2^{n}$ components.) A distinctly nonlinear
transformation ``N'' is then applied to these two qubits (we show below how
virtually any deterministic nonlinear operator can be recast into this form):%

\begin{align}
&  (a)\;\;|00\rangle+|11\rangle\longrightarrow|01\rangle+|11\rangle\nonumber\\
&  (b)\;\;|01\rangle+|10\rangle\longrightarrow|01\rangle+|11\rangle\\
&  (c)\;\;|00\rangle+|10\rangle\longrightarrow|00\rangle+|10\rangle\nonumber
\end{align}

This transformation is like an AND\ gate - it ignores the index qubit and
places the flag qubit in the state
%TCIMACRO{\TEXTsymbol{\vert}}%
%BeginExpansion
$\vert$%
%EndExpansion
1%
%TCIMACRO{\TEXTsymbol{>}}%
%BeginExpansion
$>$%
%EndExpansion
if and only if either of the original components had the state
%TCIMACRO{\TEXTsymbol{\vert}}%
%BeginExpansion
$\vert$%
%EndExpansion
1%
%TCIMACRO{\TEXTsymbol{>}}%
%BeginExpansion
$>$%
%EndExpansion
for the flag qubit.\footnote{There is one subtlety regarding nonlinear quantum
mechanics which we should address here. When the superposition principle is
abandoned, it is not immediately clear how entangled qubits will evolve. We
follow the Weinberg model, in which the time evolution for a joint system
composed of two subsystems is specified in terms of a preferred basis of
vectors for the tensor product Hilbert space. For the purpose of using the
nonlinear dynamics to perform quantum logic, we specify the joint dynamics in
terms of the basis \{%
%TCIMACRO{\TEXTsymbol{\vert}}%
%BeginExpansion
$\vert$%
%EndExpansion
b%
%TCIMACRO{\TEXTsymbol{>}}%
%BeginExpansion
$>$%
%EndExpansion
\}=\{%
%TCIMACRO{\TEXTsymbol{\vert}}%
%BeginExpansion
$\vert$%
%EndExpansion
0...00%
%TCIMACRO{\TEXTsymbol{>}}%
%BeginExpansion
$>$%
%EndExpansion
,
%TCIMACRO{\TEXTsymbol{\vert}}%
%BeginExpansion
$\vert$%
%EndExpansion
0...01%
%TCIMACRO{\TEXTsymbol{>}}%
%BeginExpansion
$>$%
%EndExpansion
, . . . ,
%TCIMACRO{\TEXTsymbol{\vert}}%
%BeginExpansion
$\vert$%
%EndExpansion
1..11%
%TCIMACRO{\TEXTsymbol{>}}%
%BeginExpansion
$>$%
%EndExpansion
\} for each subsystem. The Weinberg prescription is as follows: write the
joint state for the system $\left|  \Psi\right\rangle _{12}$ as $\sum
\limits_{b}\alpha_{b}\left|  b\right\rangle _{1}\left|  \psi_{b}\right\rangle
_{2}$ and then act on each $\left|  \psi_{b}\right\rangle _{2}$ independently
with the nonlinear transformation N.}

Step 4. The previous step is then repeated using each of the first n qubits as
the index (and the remaining n-1 qubits to define the $2^{n-1}$ subspaces).
After each iteration, the number of components in the superposition that have
a
%TCIMACRO{\TEXTsymbol{\vert}}%
%BeginExpansion
$\vert$%
%EndExpansion
1%
%TCIMACRO{\TEXTsymbol{>}}%
%BeginExpansion
$>$%
%EndExpansion
for the flag qubit doubles. After n iterations, the flag qubit is no longer
entangled with the first n qubits: it is either in the state
%TCIMACRO{\TEXTsymbol{\vert}}%
%BeginExpansion
$\vert$%
%EndExpansion
1%
%TCIMACRO{\TEXTsymbol{>}}%
%BeginExpansion
$>$%
%EndExpansion
for every component of the superposition or the state
%TCIMACRO{\TEXTsymbol{\vert}}%
%BeginExpansion
$\vert$%
%EndExpansion
0%
%TCIMACRO{\TEXTsymbol{>}}%
%BeginExpansion
$>$%
%EndExpansion
for every component of the superposition.

Step 5. Measure the flag qubit to determine the solution.

Thus, if one can perform the two qubit nonlinear transformation N one can find
the answer to an NP-complete problem with certainty in polynomial (in fact
linear) time, and using only a single evaluation of the oracle. It may be
objected that the nonlinear operator N appears arbitrary and unnatural:
indeed, it was selected exactly so as to be able to solve the stated problem.
However, the apparently arbitrary operation N can be built using ordinary
unitary operations and much simpler and more `natural' single qubit nonlinear
operators (that is, to the extent that any nonlinear operation in quantum
mechanics can be considered `natural'). One possible technique for generating
the transformation would be to use the following steps: first, act on the two
qubits with the unitary operator%

\begin{equation}
\frac{1}{\sqrt{2}}\left[
\begin{array}
[c]{llll}%
1 & 0 & 0 & 1\\
0 & 1 & 1 & 0\\
0 & 1 & -1 & 0\\
1 & 0 & 0 & -1
\end{array}
\right]
\end{equation}

This transforms the states above as follows%

\begin{align}
&  (a)\;\frac{1}{\sqrt{2}}\left[  |00\rangle+|11\rangle\right]
\longrightarrow|00\rangle\nonumber\\
&  (b)\;\frac{1}{\sqrt{2}}\left[  |01\rangle+|10\rangle\right]
\longrightarrow|01\rangle\\
&  (c)\;\frac{1}{\sqrt{2}}\left[  |00\rangle+|10\rangle\right]
\longrightarrow\frac{1}{2}\left[  |00\rangle+|01\rangle-|10\rangle
+|11\rangle\right] \nonumber
\end{align}

Next, operate on the second qubit with a simple one qubit nonlinear gate
$\widehat{n}_{-}$ that maps both
%TCIMACRO{\TEXTsymbol{\vert}}%
%BeginExpansion
$\vert$%
%EndExpansion
0%
%TCIMACRO{\TEXTsymbol{>}}%
%BeginExpansion
$>$%
%EndExpansion
and
%TCIMACRO{\TEXTsymbol{\vert}}%
%BeginExpansion
$\vert$%
%EndExpansion
1%
%TCIMACRO{\TEXTsymbol{>}}%
%BeginExpansion
$>$%
%EndExpansion
to the state
%TCIMACRO{\TEXTsymbol{\vert}}%
%BeginExpansion
$\vert$%
%EndExpansion
0%
%TCIMACRO{\TEXTsymbol{>}}%
%BeginExpansion
$>$%
%EndExpansion
. Thus%

\begin{align}
&  (a)\;\frac{1}{\sqrt{2}}\left[  |00\rangle+|11\rangle\right]
\longrightarrow|00\rangle\nonumber\\
&  (b)\;\frac{1}{\sqrt{2}}\left[  |01\rangle+|10\rangle\right]
\longrightarrow|00\rangle\\
&  (c)\;\frac{1}{\sqrt{2}}\left[  |00\rangle+|10\rangle\right]
\longrightarrow|A\rangle\nonumber
\end{align}

The third final state is unknown because we have not bothered to specify how
the non-linear gate acts on the state
%TCIMACRO{\TEXTsymbol{\vert}}%
%BeginExpansion
$\vert$%
%EndExpansion
00%
%TCIMACRO{\TEXTsymbol{>}}%
%BeginExpansion
$>$%
%EndExpansion
+
%TCIMACRO{\TEXTsymbol{\vert}}%
%BeginExpansion
$\vert$%
%EndExpansion
01%
%TCIMACRO{\TEXTsymbol{>}}%
%BeginExpansion
$>$%
%EndExpansion
-
%TCIMACRO{\TEXTsymbol{\vert}}%
%BeginExpansion
$\vert$%
%EndExpansion
10%
%TCIMACRO{\TEXTsymbol{>}}%
%BeginExpansion
$>$%
%EndExpansion
+
%TCIMACRO{\TEXTsymbol{\vert}}%
%BeginExpansion
$\vert$%
%EndExpansion
11%
%TCIMACRO{\TEXTsymbol{>}}%
%BeginExpansion
$>$%
%EndExpansion
. This omission thereby allows for flexibility in choosing the gate
$\widehat{n}_{-}$. Whatever the state
%TCIMACRO{\TEXTsymbol{\vert}}%
%BeginExpansion
$\vert$%
%EndExpansion
A%
%TCIMACRO{\TEXTsymbol{>}}%
%BeginExpansion
$>$%
%EndExpansion
may be, we can perform a unitary operation that will transform the first qubit
into the pure state
%TCIMACRO{\TEXTsymbol{\vert}}%
%BeginExpansion
$\vert$%
%EndExpansion
0%
%TCIMACRO{\TEXTsymbol{>}}%
%BeginExpansion
$>$%
%EndExpansion
while leaving the state
%TCIMACRO{\TEXTsymbol{\vert}}%
%BeginExpansion
$\vert$%
%EndExpansion
00%
%TCIMACRO{\TEXTsymbol{>}}%
%BeginExpansion
$>$%
%EndExpansion
in place. The computer is then in one of the following states%

\begin{align}
&  (a)\;|0\rangle|0\rangle\nonumber\\
&  (b)\;|0\rangle|0\rangle\\
&  (c)\;|0\rangle(x|0\rangle+y|1\rangle)\nonumber
\end{align}

A second non-linear gate $\widehat{n}_{+}$ is now required that will map the
state x%
%TCIMACRO{\TEXTsymbol{\vert}}%
%BeginExpansion
$\vert$%
%EndExpansion
0%
%TCIMACRO{\TEXTsymbol{>}}%
%BeginExpansion
$>$%
%EndExpansion
+ y%
%TCIMACRO{\TEXTsymbol{\vert}}%
%BeginExpansion
$\vert$%
%EndExpansion
1%
%TCIMACRO{\TEXTsymbol{>}}%
%BeginExpansion
$>$%
%EndExpansion
to the state
%TCIMACRO{\TEXTsymbol{\vert}}%
%BeginExpansion
$\vert$%
%EndExpansion
1%
%TCIMACRO{\TEXTsymbol{>}}%
%BeginExpansion
$>$%
%EndExpansion
(for the particular values of x and y which result from the above steps but
not necessarily for arbitrary x and y), while leaving the state
%TCIMACRO{\TEXTsymbol{\vert}}%
%BeginExpansion
$\vert$%
%EndExpansion
0%
%TCIMACRO{\TEXTsymbol{>}}%
%BeginExpansion
$>$%
%EndExpansion
unchanged. After this gate is applied, the transformation resulting from the
steps described so far is then:%

\begin{align}
&  (a)\;\frac{1}{\sqrt{2}}\left[  |00\rangle+|11\rangle\right]
\longrightarrow|00\rangle\nonumber\\
&  (b)\;\frac{1}{\sqrt{2}}\left[  |01\rangle+|10\rangle\right]
\longrightarrow|00\rangle\\
&  (c)\;\frac{1}{\sqrt{2}}\left[  |00\rangle+|10\rangle\right]
\longrightarrow|01\rangle\nonumber
\end{align}

The two qubit transformation N is then easily obtained with a NOT gate on the
second qubit and a $\pi/2$ rotation on the first qubit.\ 

Having thus shown how to generate N, the question is now reduced to that of
generating the simpler single qubit gates $\widehat{n}_{-}$ and $\widehat
{n}_{+}$. If one considers the state of a qubit as a point on the unit sphere,
then all unitary operations correspond to rotations of the sphere; and while
such rotations can place two state vectors in any particular position on the
sphere, they can never change the angle between two state vectors. A nonlinear
transformation corresponds to a stretching of the sphere, which will in
general modify this angle. The desired gates $\widehat{n}_{-}$ and
$\widehat{n}_{+}$ are two particular examples of such operations. Excepting
perhaps certain pathological cases (e.g., discontinuous transformations), it
is evident that virtually any nonlinear operator, when used repeatedly in
combination with ordinary unitary transformations (which can be used to place
the two state vectors in an arbitrary position on the sphere), can be used to
arbitrarily increase or decrease the angle between two states, as needed to
generate the gates $\widehat{n}_{-}$ and $\widehat{n}_{+}$. We describe in
detail how these gates can be obtained using the model of nonlinear quantum
mechanics put forth by Weinberg.

In Weinberg's model, the ``Hamiltonian'' is a real homogeneous non-bilinear
function $h(\psi,\psi^{*})$ of degree one, that is\cite{Weinberg 2}%

\begin{equation}
\psi_{k}\frac{\partial h}{\partial\psi_{k}}=\psi_{k}^{*}\frac{\partial
h}{\partial\psi_{k}^{*}}=h
\end{equation}

and state vectors time-evolve according to the equation%

\begin{equation}
\frac{\partial\psi_{k}}{\partial t}=-i\frac{\partial h}{\partial\psi_{k}^{*}}
\label{SE}%
\end{equation}
Following Weinberg \cite{Weinberg 2}, one can always perform a canonical
homogeneous transformation such that a two-state system (i.e., a qubit) can be
described by a Hamiltonian function%

\begin{equation}
h=n\overline{h}(a)
\end{equation}

where
\begin{align}
n  &  =|\psi_{1}|^{2}+|\psi_{2}|^{2}\\
a  &  =\frac{|\psi_{2}|^{2}}{n}%
\end{align}

It is easy to verify his solution to the time dependent nonlinear Schrodinger
equation (\ref{SE}), which is%

\begin{equation}
\psi_{k}(t)=c_{k}e^{-i\omega_{k}(a)t}%
\end{equation}

where%

\begin{align}
\omega_{1}(a)  &  =\overline{h}(a)-a\overline{h}^{\prime}(a)\\
\omega_{2}(a)  &  =\overline{h}(a)+(1-a)\overline{h}^{\prime}(a)
\end{align}

For nonlinear $\overline{h}(a)$, one sees that the frequencies depend on the
magnitude of the initial amplitude in each basis state. Intuitively, one can
imagine a transformation on the unit sphere which, instead of rotating the
sphere at a particular rate, twists the sphere in such a way so that each
point rotates at a rate which depends upon its angle $\theta$ from the axis
(clearly, this transformation involves stretching of the surface). One can
exploit this stretching of the sphere to build the gate $\widehat{n}_{-}$ as follows:

Step 1. Perform a rotation on the first qubit by an angle $\phi<45{{}^{\circ}}
$:%

\begin{align}
|0\rangle &  \longrightarrow\cos(\phi)|0\rangle-\sin(\phi)|1\rangle\\
|1\rangle &  \longrightarrow\sin(\phi)|0\rangle+\cos(\phi)|1\rangle
\end{align}

Step 2. Time-evolve the system according to the nonlinear Hamiltonian
$h=n\overline{h}(a)$. Thus%

\begin{align}
|0\rangle &  \longrightarrow\cos(\phi)|0\rangle-\sin(\phi)|1\rangle
\longrightarrow\alpha\cos(\phi)|0\rangle-\beta\sin(\phi)|1\rangle\\
|1\rangle &  \longrightarrow\sin(\phi)|0\rangle+\cos(\phi)|1\rangle
\longrightarrow\gamma\cos(\phi)|0\rangle+\delta\sin(\phi)|1\rangle
\end{align}

where $\alpha,\beta,\gamma$ and $\delta$ are phase factors. Because the
initial amplitudes of the basis states are different in the two cases, the
nonlinear Hamiltonian will cause the components to evolve at different
frequencies. As long as these frequencies are incommensurate, there is a time
t at which $\alpha$=$\gamma$=$\delta$=1 and $\beta$=-1 (to within an accuracy
$\varepsilon$). (Further, this time t is a polynomial function of the desired
accuracy $\varepsilon$.) The net result of these two steps is then
\begin{align}
|0\rangle &  \longrightarrow\cos(\phi)|0\rangle+\sin(\phi)|1\rangle\\
|1\rangle &  \longrightarrow\sin(\phi)|0\rangle+\cos(\phi)|1\rangle
\end{align}

Step 3. Reverse the first step. Thus
\begin{align}
|0\rangle &  \longrightarrow\cos(2\phi)|0\rangle+\sin(2\phi)|1\rangle\\
|1\rangle &  \longrightarrow|1\rangle
\end{align}

Essentially, we have reduced the angle between the two states by an amount
2$\phi.$ By suitable repetition of this procedure (that is, by choosing $\phi$
appropriately for each iteration), or simply by choosing $\phi$ precisely in
the first step, the states
%TCIMACRO{\TEXTsymbol{\vert}}%
%BeginExpansion
$\vert$%
%EndExpansion
0%
%TCIMACRO{\TEXTsymbol{>}}%
%BeginExpansion
$>$%
%EndExpansion
and
%TCIMACRO{\TEXTsymbol{\vert}}%
%BeginExpansion
$\vert$%
%EndExpansion
1%
%TCIMACRO{\TEXTsymbol{>}}%
%BeginExpansion
$>$%
%EndExpansion
can be mapped to within $\varepsilon$ of the state
%TCIMACRO{\TEXTsymbol{\vert}}%
%BeginExpansion
$\vert$%
%EndExpansion
0%
%TCIMACRO{\TEXTsymbol{>}}%
%BeginExpansion
$>$%
%EndExpansion
, in an amount of time which is a polynomial function of the desired accuracy.
This is the desired behavior for the nonlinear gate $\widehat{n}_{-}$. The
procedure can be modified slightly to increase the angle between state vectors
and produce the desired behavior for the gate $\widehat{n}_{+}$. We have thus
shown explicitly how to solve NP-complete problems using the Weinberg model,
using an algorithm which did not require exponentially precise operations.

To solve the problems in the class \#P, one replaces the flag qubit with a
string of $log_{2}n$ qubits and modifies the algorithm slightly - so that it
adds the number of solutions in each iteration rather than performing what is
effectively a one bit AND. In this case, a measurement of the final result
reveals the exact number of solutions.

We have demonstrated that nonlinear time evolution can in fact be exploited to
allow a quantum computer to solve NP-complete and \#P problems in polynomial
time. We have shown explicitly how to accomplish this exponential speed-up
using the Weinberg model of nonlinear quantum mechanics. In concluding, we
would like to note that we believe that quantum mechanics is in all likelihood
exactly linear, and that the above conclusions might be viewed most profitably
as further evidence that this is indeed the case. Nevertheless, the
theoretical implications and practical applications that would result from a
discovery to the contrary may warrant further investigation into the matter.

D.S.A. acknowledges support from a NDSEG fellowship, and helpful discussions
with J. Jacobson, I. Singer, S. Johnson, I. Park, and T. Wang. Portions of
this research were supported by grant \# N00014-95-1-0975 from the Office of
Naval Research, and by ARO and DARPA under grant \# DAAH04-96-1-0386 to QUIC,
the Quantum Information and Computation initiative, and by a DARPA grant to
NMRQC, the Nuclear Magnetic Resonance Quantum Computing initiative.

\pagebreak

\end{document}